\documentclass[twocolumn,showpacs,amsmath,amssymb,aps,showkeys,floatfix,a4paper]{revtex4}

\usepackage[dvips]{graphicx}
\usepackage{dcolumn}
\usepackage{bm}
\usepackage{epsfig}
\usepackage{amsfonts}
\usepackage{amssymb,amscd}

\def\lsim{\raise0.3ex\hbox{$<$\kern-0.75em\raise-1.1ex\hbox{$\sim$}}}
\def\gsim{\raise0.3ex\hbox{$>$\kern-0.75em\raise-1.1ex\hbox{$\sim$}}}

\def\odd{{I\!\!\!O}}
\def\beq{\begin{equation}}
\def\eeq{\end{equation}}
\def\bea{\begin{eqnarray}}
\def\eea{\end{eqnarray}}
\def\bq{\begin{quote}}
\def\eq{\end{quote}}

\newcommand{\rb}{\mbox{\boldmath $b$}}

\def\gappeq{\mathrel{\rlap {\raise.5ex\hbox{$>$}}
{\lower.5ex\hbox{$\sim$}}}}

\def\lappeq{\mathrel{\rlap{\raise.5ex\hbox{$<$}}
{\lower.5ex\hbox{$\sim$}}}}

\def\Toprel#1\over#2{\mathrel{\mathop{#2}\limits^{#1}}}

\begin{document}


\title{$\eta_c$ production in photon - induced interactions at the AFTER@LHC experiment as a probe of the Odderon}

\author{V.~P. Gon\c{c}alves}
\email{barros@ufpel.edu.br}
\affiliation{High and Medium Energy Group, \\
Instituto de F\'{\i}sica e Matem\'atica, Universidade Federal de Pelotas\\
Caixa Postal 354, CEP 96010-900, Pelotas, RS, Brazil}

\author{W. K. Sauter}
\email{werner.sauter@ufpel.edu.br}
\affiliation{High and Medium Energy Group, \\
Instituto de F\'{\i}sica e Matem\'atica, Universidade Federal de Pelotas\\
Caixa Postal 354, CEP 96010-900, Pelotas, RS, Brazil}

\date{\today}

\begin{abstract}

One of the open questions of the strong interaction theory is 
the existence of the Odderon, which is an unambiguous prediction of Quantum Chromodynamics, but still not confirmed  experimentally. An alternative to probe the Odderon is the   exclusive $\eta_c$ photoproduction in hadronic collisions.
As the Pomeron exchange cannot contribute to this process, its  observation  would indicate the existence of the Odderon. In this paper we estimate the $\eta_c$ production in photon - induced interactions in hadronic collisions at the AFTER@LHC experiment. We demonstrate that the experimental analysis of this process is feasible in the  AFTER@LHC experiment and that the observation of the $\eta_c$ production in nuclear collisions  is a unambiguous signature of the Odderon.

\end{abstract}


\maketitle

\section{Introduction}

The AFTER@LHC experiment opens a new kinematical regime where several questions related to the 
description of the Quantum Chromodynamics (QCD) remain without  satisfactory answers \cite{after}. 
One of these open questions is  the Odderon, which is a natural prediction of the QCD,  and determines the hadronic cross section difference between the direct and crossed channel processes at very high energies (For a review see Ref. \cite{ewerz}). 
The current experimental evidence for the Odderon is  rather scarce. 
A recent study of the data on the differential elastic $pp$ scattering shows that one needs the Odderon to describe the cross sections in the dip region \cite{dosch_ewerz} (See also Ref. \cite{lazlo,lazlo2}). The difficulties inherent in the description of $pp$ and $p\bar{p}$ collisions and the lack of  further data have made it impossible to establish the existence of the Odderon in these processes beyond reasonable doubt.

An alternative to probe the Odderon is the study of the  diffractive photoproduction of pseudoscalar mesons in hadronic collisions \cite{vic_odderon}. As the real photon emitted by one of the incident hadrons carries negative $C$ parity,  its transformation into a diffractive final state system of positive $C$ parity  requires the $t$-channel exchange of an object of negative $C$ parity. 
In perturbative QCD, the Odderon is  a $C$-odd ($C$ being the charge conjugation) compound state of three reggeized gluons, with evolution described by  the Bartels - Kwiecinski - Praszalowicz (BKP) equation \cite{bkp}, which resums terms of the order $\alpha_s(\alpha_s \log s)^n$ with arbitrary $n$ in which three gluons in a $C = -1$ state are exchanged in the $t$-channel. 
In contrast, the Pomeron corresponds to a  $C$-even  parity  compound state  of two $t$-channel reggeized  gluons, given by the solution of the Balitsky - Fadin - Kuraev - Lipatov (BFKL) equation \cite{BFKL}. 
Consequently, the Pomeron exchange cannot contribute to the production of pseudoscalar mesons and this process can only be mediated by the exchange of an Odderon. A particular promising process is the exclusive $\eta_c$ photoproduction, since  the meson mass provides a hard scale that makes a perturbative calculation possible \cite{ckms,bbcv}.

\begin{figure}[t]
\centerline{\psfig{file=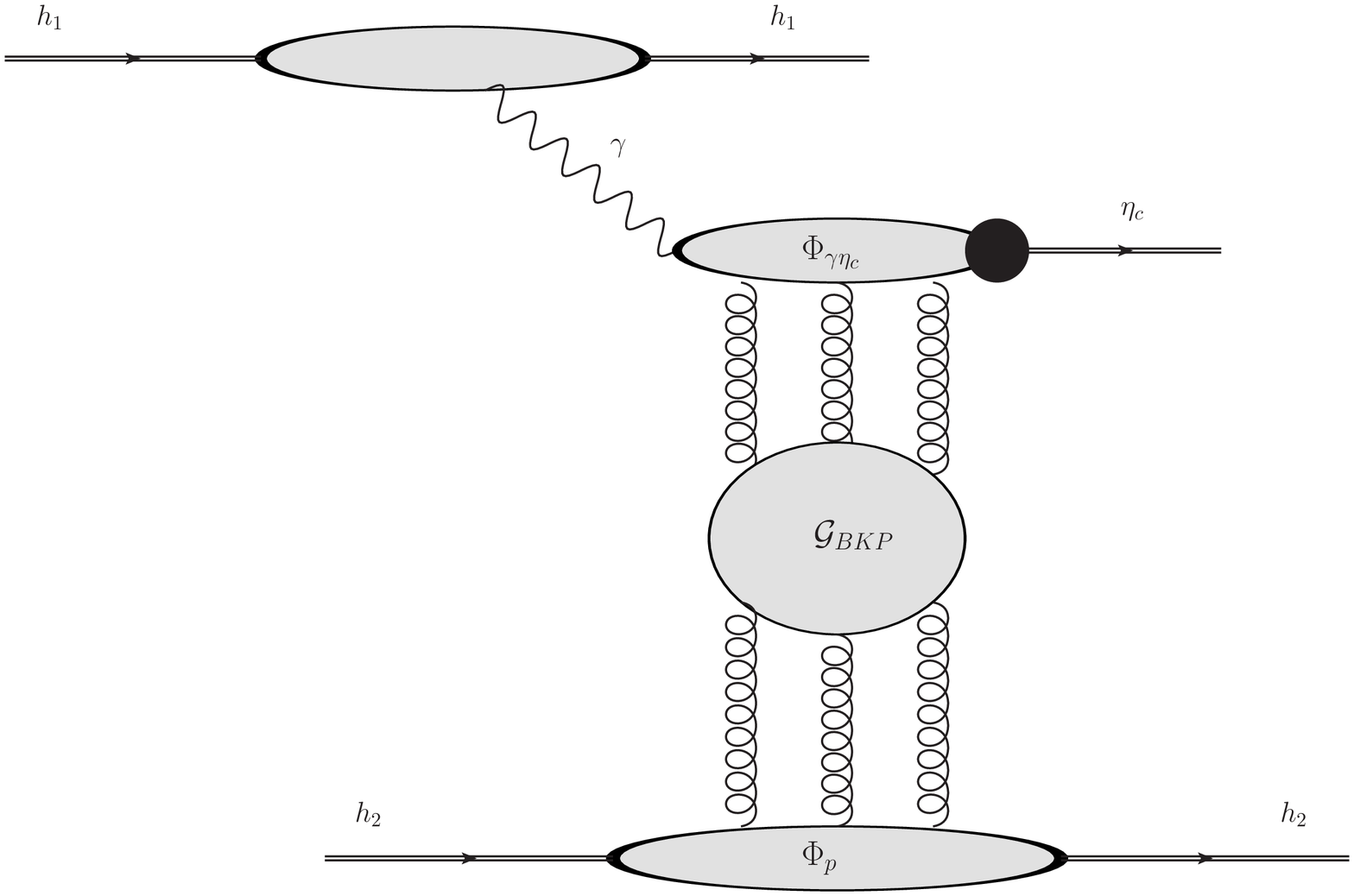,width=80mm}}
 \caption{$\eta_c$ production in photon - hadron interactions at the AFTER@LHC experiment.}
\label{fig:phothad}
\end{figure}

\begin{figure}[t]
\centerline{\psfig{file=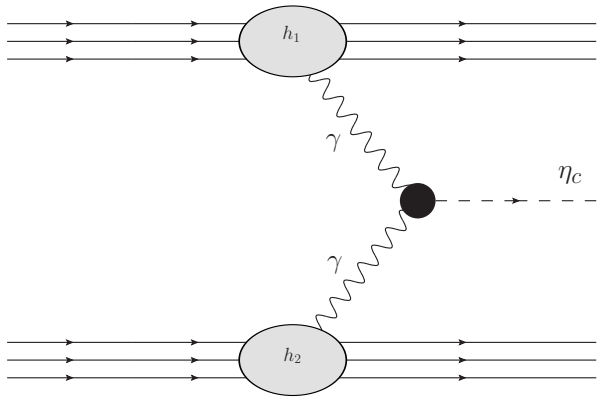,width=80mm}}
 \caption{$\eta_c$ production in photon - photon interactions at the AFTER@LHC experiment.}
\label{fig:photphot}
\end{figure}

In what follows we extend the analysis performed in Ref. \cite{vic_odderon} for the kinematical range probed by the AFTER@LHC experiment. In particular, we estimate the cross sections for the $\eta_c$ production in photon - induced interactions present in $pp, \, pA$ and $AA$ collisions. The basic idea is that in  hadron-hadron collisions at large impact parameter ($b > R_{h_1} + R_{h_2}$) and at ultra relativistic energies the electromagnetic interaction is dominant \cite{upcs}. In  heavy ion collisions, the heavy nuclei give rise to strong electromagnetic fields due to the coherent action of all protons in the nucleus, which can interact with each other. In a similar way, it also occurs when considering ultra relativistic  protons in $pp$ collisions.
The photon stemming from the electromagnetic field of one of the two colliding hadrons can interact with one photon of
the other hadron (two-photon process) or can interact directly with the other hadron (photon-hadron
process). Consequently, the $\eta_c$ can be produced in photon - hadron ($\gamma h$) and photon - photon ($\gamma \gamma$) interactions, with both processes generating two rapidity gaps in the final state. While the $\eta_c$ production in $\gamma h$ interactions represented in Fig. \ref{fig:phothad} is a direct probe of the Odderon, its production in $\gamma \gamma$ interactions (See Fig. \ref{fig:photphot}) is an important background, which should be estimated in order to separate the signal associated to the Odderon.

This paper is organized as follows. In Section \ref{coe}  we present a  brief  review of  the main concepts and formulas used  in the description of $\gamma \gamma$ and $\gamma h$  interactions in hadronic collisions and in the  exclusive $\eta_c$ photoproduction, which are required  to explain our results, which will presented in Section \ref{results}. Finally, in Section \ref{sum} we summarize our main conclusions.

\section{Photon - induced interactions in hadronic colliders}
\label{coe}

In hadronic collisions at large impact parameter  and at ultra relativistic energies  
the photon - induced cross sections for a given process can be factorized in terms of the equivalent flux of 
photons of the incident hadrons and  the  photon - photon or photon-target production cross 
section \cite{upcs}. The recent experimental results from CDF  \cite{cdf} at Tevatron, STAR \cite{star} and PHENIX \cite{phenix} at RHIC and ALICE \cite{alice,alice2} and LHCb \cite{lhcb,lhcb2} at LHC for photon-induced processes 
 in hadronic collisions  have demonstrated that a detailed analysis is feasible and that the data can be used to constrain the description of the hadronic structure  at high energies \cite{mesons} as well as to probe possible scenarios for the physics beyond the Standard Model \cite{beyond}.  These results motivate a detailed analysis of other final states in photon - induced interactions.

 Lets initially consider the $\eta_c$ production in $\gamma \gamma$ interactions.  
The cross section for the exclusive $\eta_c$ production in the two-photon fusion  process, Fig. \ref{fig:photphot},  is given by \cite{baur_jpg}
\begin{widetext}
\begin{eqnarray}
\sigma[h_1 h_2 \stackrel{(\gamma \gamma)}{\longrightarrow} h_1 \otimes \eta_c \otimes h_2] =  \int_{0}^{\infty}\! \frac{d\omega_{1}}{\omega_{1}}\! \int_{0}^{\infty}\! \frac{d\omega_{2}}{\omega_{2}}\ F(\omega_1, \omega_{2})\ \hat{\sigma}_{\gamma \gamma \rightarrow {\eta_c}}(\omega_{1}, \omega_{2})
\end{eqnarray}
\end{widetext}
where $\otimes$ represents a rapidity gap in the final state, $\omega_1$ and $\omega_2$ the energy of the photons which participate of the hard process and $\hat{\sigma}_{\gamma \gamma \rightarrow {\eta_c}}$ is the cross section for the subprocess $\gamma \gamma \rightarrow {\eta_c}$, given by
\begin{eqnarray}
\sigma_{\gamma\gamma \rightarrow \eta_c}=8\pi^{2}(2J+1)\frac{\Gamma_{\eta_c
\rightarrow\gamma\gamma}}{m_{\eta_c}}\delta(4\omega_1 \omega_2-m_{\eta_c}^{2}) \,\,,
\label{sigma-foton}
\end{eqnarray}
where $J$, $m_{\eta_c}$ and $\Gamma_{\eta_c
\rightarrow\gamma\gamma}$ are the spin, mass 
and the photon-photon partial decay width of the  $\eta_c$, respectively, and the $\delta$ function enforces energy conservation. Moreover, the function   $F$ is the folded spectra of the incoming particles (which corresponds to an ``effective luminosity'' of photons) which we assume to be given by~\cite{baur_ferreira}
\begin{widetext}
\begin{equation}
 F(\omega_1, \omega_{2}) = 2\pi \int_{R_{h_1}}^{\infty} db_{1} b_{1} \int_{R_{h_2}}^{\infty} db_{2} b_{2} \int_{0}^{2\pi} d\phi\ N_{1}(\omega_{1}, b_{1}) N_{2}(\omega_{2}, b_{2}) \Theta(b - R_{h_1} - R_{h_2}) \label{efe}
\end{equation}
\end{widetext}
where $b_i$ are the impact parameters of the hadrons in relation to the photon interaction point, $\phi$ is the angle between $\mathbf{b}_1$ and $\mathbf{b}_2$, $R_i$ are the projectile radii  and $b^2 = b_1^2 + b_2^2 - 2b_1 b_ 2 \cos \theta$. 
The theta function in Eq. (\ref{efe}) ensures that the hadrons do not overlap \cite{baur_ferreira}. The Weizs\"acker-Williams photon spectrum for a given impact parameter is given in terms of the nuclear charge form factor $F(k_{\perp}^2)$, where $k_{\perp}$ is the four-momentum of the quasi-real photon, as follows  \cite{upcs}
\begin{eqnarray}
N(\omega,\rb) = \frac{\alpha Z^2}{\pi^2 \omega}\left| \int_0^{+\infty}  dk_{\perp} k_{\perp}^2 \frac{F\left((\frac{\omega}{\gamma})^2 + \vec{b}^2\right)}{(\frac{\omega}{\gamma})^2 + \vec{b}^2} \cdot J_1(b k_{\perp}) \right|^2 \,\,,
\end{eqnarray}
where $J_1$  is the Bessel function of the first kind. For a point-like nucleus one obtains that \cite{upcs}
\begin{equation}
N(\omega,\rb) =\frac{\alpha_{em} Z^2}{\pi^2} \left(\frac{\xi}{b}\right)^2 \left\{ K_1^2(\xi) + \frac{1}{\gamma^2}  K^2_0(\xi)\right\} ,
\label{ene}
\end{equation}
with $K_{0,1}$ being the modified Bessel function of second kind, $\xi = \omega b/\gamma v$, $v$  the velocity of the hadron, $\gamma$  the Lorentz factor and $\alpha_{em}$  the electromagnetic coupling constant. This expression have been derived considering a semiclassical description of the electromagnetic interactions in peripheral collisions, which works very well for heavy ions (See e.g. \cite{baur_jpg}).  For  protons, it is more appropriate to obtain the equivalent photon spectrum  from its elastic form factors in the dipole approximation (See e.g. \cite{david}). An alternative  is to use Eq. (\ref{ene}) assuming $R_p = 0.7$ fm for the proton radius, which implies a  good agreement with the parametrization of the luminosity obtained in \cite{Ohnemus:1993qw} for proton-proton collisions (For a more detailed discussion see Ref. \cite{vicwerdaniel}).

In the case of the $\eta_c$ production in photon - hadron interactions the total cross section is given by
\begin{eqnarray}
\sigma [h_1 h_2 \stackrel{(\gamma h)}{\longrightarrow} h_1 \otimes \eta_c \otimes h_2] =  \sum_{i=1,2} \int dY \frac{d\sigma_i}{dY}\,,
\label{sighh}
\end{eqnarray}
where  ${d\sigma_i}/{dY}$ is the rapidity distribution for the photon-target interaction induced by the hadron $h_i$, which can be expressed as 
\begin{eqnarray}
\frac{d\sigma_i}{dY} = \omega n_{\gamma/h_i}(\omega)\,\cdot\,\sigma_{\gamma h_j \rightarrow \eta_c h_j} (W_{\gamma h_j}^2) \,\,\,\,\,\,(i\neq j)\,.
\label{rapdis}
\end{eqnarray}
where $W_{\gamma h}^2=2\,\omega \sqrt{s_{\mathrm{NN}}}$  and ${s_{\mathrm{NN}}}$ are  the  c.m.s energy squared of the
photon - hadron and hadron-hadron system, respectively.
Moreover, $n_{\gamma/h_i}(\omega)$ is the $\rb$-integrated photon flux associated to the hadron $h_i$, which can be obtained considering the requirement that  the photon - induced processes are  not accompanied by hadronic interaction (ultra-peripheral
collision). An analytic approximation for the equivalent photon flux of a nuclei can be calculated, which is given by \cite{upcs}
\begin{eqnarray}
n_{\gamma / A}(\omega) & = &  \int_{b_{min}} d^2\rb \, N(\omega,\rb) \nonumber \\
& = & \frac{2\,Z^2\alpha_{em}}{\pi\,\omega}\, \left[\bar{\eta}\,K_0\,(\bar{\eta})\, K_1\,(\bar{\eta})+ \frac{\bar{\eta}^2}{2}\,{\cal{U}}(\bar{\eta}) \right]\,
\label{fluxint}
\end{eqnarray}
where   $\bar{\eta}=\omega\, b_{min} /\gamma_L$ (with $\gamma_L$ being the Lorentz boost  of a single beam), $b_{min} = R_{h_1} + R_{h_2}$ and  ${\cal{U}}(\bar{\eta}) = K_1^2\,(\bar{\eta})-  K_0^2\,(\bar{\eta})$. On the other hand, for   proton-proton collisions, we assume that the  photon spectrum of a relativistic proton is given by  \cite{Dress},
\begin{widetext}
\begin{eqnarray}
n_{\gamma/p}(\omega) =  \frac{\alpha_{\mathrm{em}}}{2 \pi\, \omega} \left[ 1 + \left(1 -
\frac{2\,\omega}{\sqrt{S_{NN}}}\right)^2 \right] .
\left( \ln{\Omega} - \frac{11}{6} + \frac{3}{\Omega}  - \frac{3}{2 \,\Omega^2} + \frac{1}{3 \,\Omega^3} \right) \,,
\label{eq:photon_spectrum}
\end{eqnarray}
\end{widetext}
where $\Omega = 1 + [\,(0.71 \,\mathrm{GeV}^2)/Q_{\mathrm{min}}^2\,]$ and $Q_{\mathrm{min}}^2= \omega^2/[\,\gamma_L^2 \,(1-2\,\omega /\sqrt{s_{NN}})\,] \approx (\omega/
\gamma_L)^2$.

The exclusive $\eta_c$ photoproduction, which is the main input in our calculations [See Eq. (\ref{rapdis})] can be obtained using the impact  factor representation, proposed by Cheng and Wu \cite{ChengWu} many years ago. In this representation, the amplitude for a large-$s$ hard collision process can be factorized in {three parts}: the two impact factors of the colliding particles and the Green's function for the  three interacting reggeized gluons, which is determined by the BKP equation and is  represented by ${\cal G}_\mathrm{BKP}$ hereafter. 
The differential cross section for the process $\gamma + h \rightarrow \eta_c + h$  is given by \cite{bbcv}
\begin{eqnarray}
\frac{d\sigma}{dt} = \frac{1}{32 \pi}\sum_{i=1,2} |{\cal{A}}^i|^2 \,\,, 
\end{eqnarray}
where ${\cal{A}}^i$ is the amplitude for a given transverse polarization $i$ of the photon, which can be expressed as a convolution of the impact factors for the proton ($\Phi_p$) and for the $\gamma \eta_c$ transition ($\Phi^i_{\gamma \eta_c}$) with the Odderon Green function:
\begin{eqnarray}
{\cal{A}}^i = \frac{5}{1152} \frac{1}{(2\pi)^8} \langle \Phi^i_{\gamma \eta_c}|{{\cal{G}}_{BKP}}| \Phi_p \rangle\,\,.
\end{eqnarray}
Differently from $\Phi^i_{\gamma \eta_c}$, that can be calculated perturbatively \cite{ckms}, the impact factor $\Phi_p$ that describes the coupling of the Odderon to the proton is non-perturbative and should be modelled. 
In our calculations we consider the model	 used in  Refs. \cite{ckms,bbcv}. Moreover, we assume that the  Odderon Green function ${\cal{G}}_{BKP}$ is described in terms of the solution of the BKP equation \cite{bkp}, with the energy dependence being determined by the Odderon intercept $\alpha_{\odd}$. In particular, we consider the solution obtained by 
Bartels, Lipatov and Vacca (BLV) \cite{blv} that have found a solution for the BKP equation with intercept $\alpha_{\odd}$ exactly equal to one. For comparison we also consider the solution obtained by Kwiecinski and collaborators in Ref. \cite{ckms} (CKMS model hereafter), which has considered a simplified three gluon exchange model for the Odderon that implies an energy independent cross section. In our calculations we will use  a realistic value for $\alpha_s$ (= 0.3).

\section{Results}
\label{results}

In what follows we present our predictions for the $\eta_c$ production in $\gamma h$ and $\gamma \gamma$ interactions considering the kinematical range which will be probed by the AFTER@LHC experiment. Basically, we assume $\sqrt{s_{NN}} = 115/\,72/\,72$ GeV for $pp/Pbp/PbPb$ collisions, which implies that $\sqrt{s_{\gamma h}} \le 44/\, 12/\, 9$ GeV, respectively. Similarly, it is possible to obtain that $\sqrt{s_{\gamma \gamma}} \le 17/\,2.0/\,1.0$ GeV. Consequently, the $\eta_c$ production in $\gamma \gamma$ interactions only is present in $pp$ collisions.  
In other words, the measurement of the exclusive $\eta_c$ production in $Pbp$ and $PbPb$ collisions can be considered a direct probe of the Odderon.  Moreover, in our calculations we take into account that the typical rapidity range which is expected to be reachable by the AFTER@LHC experiment is $-3.0 \le Y_{c.m.} \le 0.5$.

In Table \ref{I} we present our predictions for the  total cross sections. We predict cross sections for the $\eta_c$ production in $Pbp$ and $PbPb$ collisions that are a factor $\geq 10^4$ larger than the $pp$ predictions. This enhancement is directly associated to the nuclear photon flux and the nuclear dependence of the photon - hadron cross section.
As the photon flux is proportional to $Z^2$, {because} the electromagnetic field surrounding the ion is very larger than the proton one due to the coherent action of all protons in the nucleus,  the $Pbp$ and $PbPb$ cross sections are amplified by this factor. Moreover, our predictions for the $\eta_c$ production in $PbPb$ collisions also are amplified by the mass number $A$, since in  our calculations for the nuclear case we are assuming in a first approximation that $\sigma (\gamma A \rightarrow \eta_c A) = A . \sigma (\gamma p \rightarrow \eta_c p)$. For the  exclusive $\eta_c$ production in $pp$ collisions we predict values of the order of a fraction of pb, with the BBVC prediction being a factor of $\approx 6$ larger than the CKMS one. This  enhancement is directly associated to the energy dependence present in the BBVC model, which implies that the $\gamma h$ cross section increases at smaller energies, while the CKMS predicts an energy independent cross section. For the $\eta_c$ production in $Pbp$ and $PbPb$ collisions, we predict cross sections of the order of nb for the exclusive $\eta_c$ photoproduction in $PbPb$ collisions at AFTER@LHC experiment. Moreover, we predict  that the $Pbp$ cross sections are two orders of magnitude smaller than those predicted for $PbPb$ collisions.

Lets now estimate the background associated to the $\eta_c$ production in $\gamma \gamma$ interactions for $pp$ collisions. 
Assuming that the photon spectrum for the proton is given by Eq. (\ref{ene}), with $R_p = 0.7$ fm,   $m_{\eta_c}  = 2.983$ GeV and $\Gamma (\eta_c \rightarrow \gamma \gamma) = 5.0$ keV, we predict that $\sigma[p p \stackrel{(\gamma \gamma)}{\longrightarrow} p \otimes \eta_c \otimes p] = 2.2$ pb, which is  a factor $\gtrsim 8$ larger than the predictions for the $\eta_c$ production in photon - hadron interactions. As both processes generate two rapidity gaps in the final state, the detection of the gaps is not, in a first analysis, an efficient trigger for the separation of the $\gamma h$ production of the $\eta_c$. An alternative is the reconstruction of the  entire event with a cut  on the summed transverse momentum of the event \cite{david}.  As the typical photon virtualities are very small, the hadron scattering angles are very low. Consequently, we expect that  a different transverse momentum distribution of the scattered hadron, with $\gamma h$ interactions predicting larger $p_T$ values. 
 In contrast, the background is not present in nuclear collisions, since the 
maximum $\gamma \gamma$ center-of-mass energies in $Pbp$ and $PbPb$ collisions are smaller than threshold of production.

Considering the {design} luminosities at AFTER@LHC for $pp$  (${\cal L}_{\mathrm{pp}} = 2 \times 10^{4}$ pb$^{-1}$yr$^{-1}$), $Pbp$ (${\cal L}_{\mathrm{Pbp}} = 1.1$ pb$^{-1}$yr$^{-1}$) and $PbPb$ collisions (${\cal L}_{\mathrm{PbPb}} = 7.0 \times 10^{-3}$ pb$^{-1}$yr$^{-1}$) we can calculate the production rates (See Table \ref{I}). 
Although the cross section for the exclusive $\eta_c$ photoproduction  in $PbPb$ collisions is much larger than in $pp$ collisions, the event rates are higher in the $pp$ mode  due to its larger luminosity. In particular, we predict that the events rate/year for $pp$ collisions at $\sqrt{s} = 115$ GeV should be larger than 1000. On the other hand, for $Pbp$ and $PbPb$ collisions at $\sqrt{s} = 72$ GeV  we predict that the events rate/year should be larger than 30. Although smaller than the $pp$ predictions, the observation of the $\eta_c$ production in nuclear collisions  would clearly indicate the existence of the Odderon.

\begin{table}[t]
\begin{center}
\begin{tabular}{||c|c|c||}
\hline 
$h_1 h_2$ & CKMS & BBCV \\
\hline
\hline 
$pp$ ($\sqrt{s} = 115$ GeV) & $0.05$ pb (1000.0) & $0.30$ pb (6000.0) \tabularnewline
\hline 
$Pbp$ ($\sqrt{s} = 72$ GeV) & $28.1$ pb (31.0) & $356.6$ pb (393.0) \tabularnewline
\hline
$PbPb$ ($\sqrt{s} = 72$ GeV) & $5870.0$ pb (41.0) & $74366.0$ pb (520.0) \tabularnewline
\hline
\hline
\end{tabular}
\caption{Cross sections (event rates/year) for the exclusive $\eta_c$ photoproduction in $pp/Pbp/PbPb$ collisions at AFTER@LHC experiment. }
\label{I}
\end{center}
\end{table}

\section{Summary}
\label{sum}

In the last years, the physics of the Odderon has become an increasingly active subject of research, both from theoretical and experimental  points of view. On the theoretical side, the investigation of the Odderon in pQCD has led to discovery of relations of high energy QCD to the theory of integrable models \cite{korchemsky} and two leading solutions of the BKP evolution equation were obtained \cite{janik,blv}, with the intercept being close to  or exactly one, depending on the scattering process (See also Refs. \cite{outros_odderon}). In contrast,
 on the experimental side, the  current  evidence for the Odderon is very unsatisfactory. 

 In Ref. \cite{vic_odderon} we have proposed the study of the exclusive $\eta_c$ production in hadronic collisions at LHC energies as a probe of the Odderon (For other possibilities see Ref. \cite{pheno_odderon}). In this paper we extend that previous analysis for the kinematical region which would be probed by the AFTER@LHC experiment. As the exclusive $\eta_c$ photoproduction is only possible if the Odderon is exchanged between the vector meson and the hadron, the observation of such processes would clearly indicate the existence of the Odderon. 
We have estimated the $\eta_c$ cross section considering photon - hadron interactions in $pp/Pbp/PbPb$ collisions. Moreover, the background associated to the $\eta_c$ production by $\gamma \gamma$ interactions was calculated. We have that the background is only present in $pp$ collisions, which makes the observation of the exclusive $\eta_c$ production in $Pbp$ and $PbPb$ a signature of the  Odderon. We predict  total cross sections of order of pb (nb) for $pp \, (PbPb)$ collisions and large values for the event rates/year, which makes, in principle, the experimental analysis of this process feasible at AFTER@LHC experiment. 

\section*{Acknowledgements}
 This work was partially financed by the Brazilian funding agencies CNPq, CAPES and FAPERGS.



\end{document}